\documentstyle[12pt]{article}
\begin{document}

\begin{titlepage}

\pagestyle{empty}

\begin{flushright}
{\footnotesize Brown-HET-991

March 1995}
\end{flushright}

\vskip 1.0cm

\begin{center}
{\Large \bf FRW Type Cosmologies with Adiabatic Matter Creation}

\vskip 1cm

{J. A. S. Lima$^{1, 2}$, A. S. M. Germano$^2$ and L. R. W. Abramo$^1$}

\end{center}

\vskip 0.5cm

\begin{quote}
{\small $^1$ Physics Department, Brown University,
Providence, RI 02912,USA.

$^2$ Departamento de F\'{\i}sica Te\'orica e Experimental,
     Universidade \\
$^{ }$ $^{ }$ Federal do Rio Grande do Norte,
     59072 - 970, Natal, RN, Brazil.}
\end{quote}

\vskip 3.5cm

\begin{abstract}

\noindent Some properties of cosmological models with matter creation are
investigated
in the framework of the Friedman-Robertson-Walker (FRW) line element. For
adiabatic matter creation, as developed by Prigogine and coworkers, we
derive a simple expression relating the particle number density $n$ and
energy density $\rho$ which holds regardless of the matter creation rate.
The conditions to generate inflation are discussed and
by considering the natural phenomenological matter
creation rate
$\psi =3 \beta nH$, where $\beta$ is a pure number of the order
of unity and $H$ is the Hubble parameter, a minimally modified hot
big-bang model is proposed. The dynamic properties of
such models can be
deduced from the standard ones simply by replacing the adiabatic
index $\gamma$ of the equation of state by an effective parameter
$\gamma_{*} = \gamma (1 - \beta)$. The thermodynamic
behavior
is determined and it is also shown
that ages large enough to agree with observations are obtained even
given the high values of $H$ suggested by recent measurements.

\end{abstract}

\end{titlepage}

\pagebreak

\section{Introduction}

\hspace{.3in} The origin of the material content (matter plus radiation)
filling the presently
observed universe remains one of the most
fascinating unsolved mysteries in cosmology even though
many authors worked out to understand the
matter creation process and its effects
on the evolution of the universe [1-27].

Radiation and matter constituents can quantum-mechanically be
produced in the context of Einstein's theory or, more
generally, in any relativistic theory of gravitation. Such a
process has been systematically
investigated by Parker and coworkers\cite{Parker1}
by considering the Bugoliubov mode-mixing technique
in quantum field theory. This approach, roughly speaking, follows
naturally from the fact that in curved spacetimes, as well as in
accelerated frames, it is usually impossible to
fix {\it a priori} a unique vacuum state for quantum
fields\cite{Birrell}. In particular, this
means that an observer with a detector will detect at late times a
nonvanishing flux of particles in a state initially set
up to be empty of
particles (Fulling-Unruh-Davies effect). Unfortunately, due to
expected back-reaction effects, it
is not so clear that such a mechanism can account for sufficient particle
creation to explain either the cosmic background radiation or
the matter content of the universe.

An alternative approach to matter creation, the development
of which gave rise to deep physical insights for
theoretical cosmology, was suggested by Tryon\cite{Tryon} and
independently by Fomin\cite{Fomin}. They argued that if the net value of all
conserved quantities of the universe is zero, as for instance,
the total energy (gravitational plus material),
then a universe whose duration is quantically restricted by the uncertainty
relation $\Delta E \Delta t \geq \hbar /2$, could have emerged as a
vacuum fluctuation. No specific
scenario was proposed by these authors however, in such an approach, the
universe
should be spatially closed in order
to have all net charges identically zero. These ideas guided
Zeldovich and other researchers to investigate the possibility
that the classical space-time came into existence from a quantum-gravitational
tunneling process termed ``spontaneous birth of the
universe''\cite{Zeldovich,Vilenkin}.

A different but somewhat related line of development was pursued by Brout,
Englert and Gunzig\cite{BroutEG}. They proposed a concrete scenario that
provided a simultaneous generation of matter and curvature from a quantum
fluctuation of the Minkowski spacetime vacuum. In this model, after a
first stage of creation, the universe enters a de
Sitter phase by which some cosmological
problems (horizon, flatness, etc.) are solved. In a
subsequent period, the system finally achieves the standard FRW phase.

Many attempts to treat the matter creation process at a phenomenological
macroscopic level have also long been considered
in the literature based on rather disparate
motivations (for a review of early literature see Ref.\cite{Narlikar}).
There have also been some claims [9-11], that particle creation during or
near the Planck era could classically be modeled by bulk viscosity
stress(second viscosity). This is an interesting
connection since irreversible processes are believed to play a fundamental
role in the problem of time-asymmetry\cite{Hu2}. In this case, the
usual thermodynamic ``arrow of time", translated in this context as
entropy generation due to matter creation,
could provide a natural explanation of the
``arrow of time" in the cosmological domain.
The enlargement of the traditional FRW equilibrium equations to
include these effects has also, at least, two additional goals, namely:
to explain the observed high entropy of the cosmic background radiation and
to avoid the initial singular state existing in the
standard equations [13-16].

More recently, irreversible processes have become the subject of study once
again in connection with inflationary universe
scenarios [17-20].
The basic idea is that bulk viscosity (matter creation) contributes
at the level of the Einstein field equations (EFE) as a negative pressure term.
It turns out that effective negative pressures are the key
condition to generate inflation. Specifically, Barrow\cite{Barrow}
introduced this idea in the framework of the new
inflationary scenario. He claimed that
particle creation due to nonadiabatic decay of
the field driving the slow-rollover inflation can macroscopically be
described by the viscous cosmological model found by
Murphy \cite{Murphy}, for which the bulk viscosity coefficient is
proportional to the energy density of the fluid.

The above considerations show that since the very beginning bulk
viscosity has been widely interpreted as a phenomenological description
of the matter
creation process in the cosmic fluid(see Refs. quoted in \cite{Johi} for
recent papers in this
line). However, regardless of this macroscopic analogy as well as any
microscopic description, it is important in itself to know how matter
creation  can be incorporated in the classical Einstein field equations.
This question  was seriously considered in the pioneering article of
Prigogine and coworkers
\cite{Prigogine}, who implicitly pointed out that the bulk viscosity and
matter creation are not only independent processes but, in general, lead to
different histories of the universe evolution. They argued that, at the
expense of the gravitational field, matter creation can occur
only as an irreversible process constrained by the usual requirements
of nonequilibrium thermodynamics. The crucial ingredient of the new
approach is the explicit use of a balance equation for the number of
created particles in addition to the Einstein field equations.
When properly combined with the thermodynamic second law, such an
equation leads naturally to a reinterpretation of the stress tensor
corresponding to an additional
negative pressure term which, as should be expected, depends on the
matter creation rate. This is in marked contrast to the bulk viscosity
formulation, in which entropy is produced but, the number particle
conservation law is taken for granted. These results were further
discussed and generalized
by Calv\~{a}o, Lima and Waga\cite{L.C.W.,C.L.W.} through a
covariant formulation allowing specific entropy
variation as usually expected for nonequilibrium processes in fluids. The
issue of why the processes of bulk viscosity and matter
creation are not equivalent either from a dynamic or
a thermodynamic point of view has been recently discussed in the
literature \cite{L.G.,Gabriel}.

The macroscopic irreversible approach to matter creation has also been
applied by these authors
to early universe physics. For instance, Prigogine
et al\cite{Prigogine} obtained a scenario
quite similar to
that proposed in Ref.\cite{BroutEG}, in which the universe emerges
from an
initial Minkowski vacuum. Thus, to a certain extent, this work can be
viewed as the macroscopic counterpart of the ideas originally
proposed by Tryon\cite{Tryon} and Fomin\cite{Fomin} and further
semiclassically developed  in the
model proposed by Brout et al.\cite{BroutEG}. However, unlike the
reversible semiclassical
equations considered in the latter, the phenomenological approach
provides, in a natural way, the entropy burst accompanying
the production of matter.

In this article we focus our attention on the ``adiabatic'' matter creation
as originally formulated in Ref. \cite{Prigogine} and somewhat
clarified in \cite{C.L.W.}.
As we shall see, unlike the standard model, to construct a definite
scenario with matter creation one needs to solve a system of three
coupled differential equations since the balance equation for the number
density has been added to the pair of independent EFE.
In principle, the full integration of such a system is not a trivial
task because it depends on the somewhat unknown matter creation rate. However,
as will be seen, it is possible to establish a simple
relation between the equations for the particle
number and energy density which holds regardless of
the matter creation rate. This result will allow us
to write the differential equation for the scale
factor in terms only of the matter creation rate,
thereby simplifying the analysis of the physically
admissible models as well as their comparison
with the bulk viscous universes. The conditions to
generate inflation in this context will be generically discussed,
however, unlike previous work on this
subject [22-27], we are more
interested in the late stages of universe evolution. In this sense, a
new class of cosmologies endowed with matter creation, leading to definite
predictions in the present phase, is proposed. As argued,
for all values of the curvature parameter, this is
the simplest class of hot big-bang cosmologies driven by the matter creation
process. As in the standard model,
the thermodynamic behavior is readily computed and it is
also shown that ages large enough
to agree with observation can be obtained even given the high values of
the Hubble parameter suggested by the recent measurements
\cite{Pierce,Freedman}.

\section{FRW Equations With Matter Creation}

\hspace{.3in} Let us now consider the FRW line element $(c=1)$

\begin{equation}
  ds^2 = dt^2 - R^{2}(t) (\frac{dr^2}{1-k r^2} + r^2 d \theta^2 + r^2
                    sin^{2}(\theta) d \phi^2)
      \quad ,
\end{equation}
where $k=0$, $\pm 1$ is the curvature parameter of the spatial section
and $R$ is the scale factor.

     In that background, the nontrivial EFE for a fluid endowed with
matter creation and the balance equation for the particle number
density can be written as [23-25]

\begin{equation}
    8\pi G \rho = 3 \frac{\dot{R}^2}{R^2} + 3 \frac{k}{R^2} \quad ,
\end{equation}

\begin{equation}
   8\pi G (p+p_{c}) = -2 \frac{\ddot{R}}{R} - \frac{\dot{R}^2}{R^2} -
    \frac{k}{R^2}
                       \quad ,
\end{equation}

\begin{equation}
      \frac{\dot{n}}{n} + 3 \frac{\dot{R}}{R} =
           \frac{\psi}{n}
               \quad ,
\end{equation}
where an overdot means time derivative and $\rho$, $p$, $n$ and
$\psi$ are the energy density, thermostatic pressure, particle number
density and matter creation rate, respectively. The creation pressure
$p_{c}$ depends on the matter creation rate, thereby coupling Eqs.
(3) and (4) to each other and, although indirectly, both of them with (2)
through
of the energy conservation law which is contained in the EFE themselves.
For ``adiabatic'' matter creation, this pressure assumes the following form
(See Ref. \cite{C.L.W.} for a more general expression)

\begin{equation}
    p_{c} = - \frac{\rho + p}{3nH} \psi ,
\end{equation}
where $H = {\dot {R}}/R$ is the Hubble parameter.

     As it stands, the above system is underdetermined since there are
six unknowns, namely: $\rho$, $p$, $p_c$, $n$, $R$, $\psi$  and only
three equations plus the constraint (5). It thus follows that one needs
to provide two more relations in order to construct a definite cosmological
scenario with matter
creation. The first constraint takes the form of an
equation of state which is supplied by thermodynamical considerations.
The one usually employed in cosmology is the so-called ``gamma-law''
equation of state

\begin{equation}
              p = (\gamma - 1)\rho
    \quad ,
\end{equation}
where the constant $\gamma$ lies in the interval [0,2].
The second constraint is a specification of
the exact form of the matter creation $\psi$ (which
should be determined from a more fundamental theory involving quantum
processes). At this point, the procedure followed in the literature has
been: (a) to integrate Eq.(4) assuming a given phenomenological law for
$\psi$ (b) to insert the expression of $n$ into (5) and using (6) to
obtain the evolution equation  for the scale factor.
Here, we consider a more general and somewhat more comprehensive approach.
Firstly, we will visualize the kind of coupling existing among the balance
equation (4) and the EFE. To that end, we establish the differential
equation for $R$ as a function of $\psi$ and $n$.
Combining Eqs.(2)
and (3) with (5) and (6) it follows that
\begin{equation}
     R\ddot{R} + (\frac{3 \gamma - 2}{2} - \frac{\gamma
     \psi}{2nH}) \dot{R}^2 + ( \frac{3 \gamma - 2}{2} -
     \frac{\gamma \psi}{2nH}) k = 0                     \quad.
\end{equation}
This is a very enlightening expression in determining
the
effects of $ \psi $ on the evolution of the
scale factor.
Since the $3nH$ term in (4) measures the
variation of $n$ only due to the expansion of the universe, it proves
convenient to introduce the dimensionless
and in general time-dependent parameter
\begin{equation}
    \beta = \frac{\psi}{3nH}
       \quad,
\end{equation}
in order to measure the effects of the matter creation
rate. As
expected, for $\psi = \beta$ = 0, Eq.(7)
reproduces the general FRW
equation, thereby decoupling the subsystem formed by
equations (2) and (3) from
(4). Indeed, unlike claims by the authors of \cite{Prigogine}, such
decoupling for
$\psi = 0$  happens only if the
``$\gamma$ - law'' has been adopted (Ref. \cite{C.L.}
examined the coupling of
(2) - (4) for $\psi = 0$ and a more general equation
of state).
 If $\psi$ is different from zero but $\psi << 3nH$,
 that is, $ \beta << 1$, the effects of $\psi$ may be safely neglected.
Physically, one may expect that the most interesting solutions of (7)
arise during the phases in which the parameter
$\beta$ is of the order of unity.

      To proceed further, we now establish a general
expression relating $n$ and $\rho$. By inserting (5) and
using (6) in the energy conservation law

\begin{equation}
           \dot{\rho} + 3({\rho} + p + p_{c})H = 0
 \quad,
\end{equation}
it takes the form

\begin{equation}
      \dot{\rho} + 3 \gamma \rho{H} =
       \gamma \rho \frac{\psi}{n}
  \quad.
\end{equation}
Therefore, comparing (10) with (4) it follows that

\begin{equation}
  {\gamma} \frac{\dot{n}}{n} =  \frac{ \dot{\rho}}{\rho}           \quad,
\end{equation}
the solution of which is

\begin{equation}
n =  n_{o}{(\frac{\rho}{\rho_{o}})}^{\frac{1}{\gamma}}        \quad,
\end{equation}
where $n_{o}$ and $\rho_{o}$ are the values of $n$ and
$\rho$ at a given instant (from now on the index o denotes the
present values of the parameters). It is worth mentioning
that (12) holds regardless of the specific form assumed for the matter
creation rate $\psi$. It reduces to the right limit for a dust filled
universe since in this case
$\rho = nM$, where $M$ is the mass of the created
particles. It is also easy to see that (12) does not remain valid in the
more general formulation proposed in Ref. \cite{C.L.W.}.

      In summary, the set of equations (2)-(6) has been reduced to
Eq.(7) together with relation (12). Thus, taking into
account relation (2) for the energy density $\rho$, the
integration of the evolution equation for the scale
factor depends only on the form of $\psi$. In principle,
since matter creation is essentially a quantum process, the
corresponding rates should be obtained from a quantum field
theory in the presence of gravitational fields. However,
in the absence of a well accepted model, the natural way
is to investigate physically interesting solutions
of (7) by adopting a phenomenological description. In the
case of viscous models, for instance, the analogous step
is to to prescribe a form of the bulk viscosity coefficient. The one
widely adopted in the literature is
$\xi = \alpha{ \rho}^{\nu}$,  where   $ \alpha$  is a
dimensional constant and $\nu$  lies in  interval  [0,1] (see Ref.
\cite{Belinski}).

For matter creation, Prigogine et al. \cite{Prigogine} examined the
consequences
of assuming a rate $\psi = \alpha{H^2}$ ($\alpha$ constant)  for a dust filled
$FRW$
flat model. As one  can see from (12) and (2), such a choice corresponds to
$\psi = \alpha 8\pi G \rho /3$, that is,  the same type of phenomenological
expression first considered by Murphy \cite{Murphy}
in the context of a cosmology with viscosity.
However, as discussed in \cite{L.G.}, the models are quite different from
a physical point of view.
Of course, if one chooses $\psi = \psi(n)$, Eq.(12) can always be used
to rewrite the relation in terms of
$\rho$. For instance, in the case considered in
Ref.[26],  $ \psi $  proportional to $ n{H^2} $ implies that
 $ \psi $ scales with ${\rho}^{1+ \frac{1}{\gamma}}$.

\section{Inflation and Matter Creation}

      As is well known, inflation is a
theory of the early universe based not only on the
possible existence of a primordial scalar field.
To obtain the required dynamics, the potential of such a field needs to
be able to, at least for a finite period,
generate a state of negative stress, which
is the key
condition to realize inflation.
As shown by Guth and Sher\cite{Guth-Sher}, a  prerequisite for
inflation to work is a departure from
thermodynamic equilibrium.
In this context, it is naturally of interest to establish,
at least
qualitatively,
how inflation can be implemented in the present irreversible matter
creation theory. In
order to analyze this issue, we first define
an effective ``adiabatic index''
$\gamma_{*} = ( 1 -  \frac{\psi}{3nH})\gamma$, so that Eq.(7) assumes the
following FRW type form :

\begin{equation}
       R\ddot{R} + (\frac{3\gamma_{*} - 2}{2}){\dot{R}^2}
      + (\frac{3\gamma_{*} - 2}{2}) k = 0
 \quad,
\end{equation}
which can be obtained using only the EFE (2) and (3) together
with the effective
equation of state $p = ( \gamma_{*} -1 )\rho$. As
usual, for
early times,  we will neglect the spatial curvature contributions.
In this case, Eq.(9) can be rewritten as

\begin{equation}
   \dot{H} + \frac{3 \gamma_{*}}{2}{H^2} = 0
  \quad.
\end{equation}
Hence, from the expression for $\gamma_{*}$  we see
that the condition for
exponential inflation $(\dot{H} = 0)$ is given by

\begin{equation}
        \psi = {3nH}
\quad,
\end{equation}
or equivalently (from (8)), $\beta = 1$. Physically this is not a
surprising fact, since the matter creation rate given by (15) has
exactly the value that compensates for the dilution of particles due to
expansion.
As we shall see in section 5, exponential inflation occurs isothermically
so that there is not extreme supercooling and
violent subsequent reheating, as happens in all variants of inflation
driven by a scalar field.
As a matter of a fact, with continuous matter creation, even in the power-law
inflationary case, the decrease in temperature is much less
than in the adiabatic case. The physical reason is quite
simple. In this context, the entropy generation is concomitant with
inflation, differently of what happens in the usual inflationary variants,
where entropy is generated after inflation by a highly
nonadiabatic process. In the new inflationary scenario,
for instance, the temperature should decrease during the
slow-rollover phase at least by a factor of ${10}^{-28}$ in order
to maintain the radiation entropy constant. In the
present scenario, the matter creation continuously reheats
the medium so that the temperature need not decrease so
drastically. Essentially, this is the same result
arising in the framework of inflationary models driven by bulk
viscosity (see for instance Ref\cite{A.P.W.}).

Now, recalling
that a violation of the strong energy condition
$(\gamma_{*}< \frac{2}{3})$ is a sufficient
condition for ``power law"
inflation, one can show that (15), in this case,
must be replaced by

\begin{equation}
       \psi > ( 1 - \frac{2}{3\gamma}){3nH}
  \quad,
\end{equation}
or $\beta > 1 - \frac{2}{3\gamma}$. Note that
due to the matter
creation process, either exponential or ``power law'' inflation are
now compatible with the existence of usual matter
described by the
``$\gamma$-law'' equation of state.
In fact, Eqs. (15) and (16) do not impose any constraint on the $\gamma$
parameter. In this way, one may now refer, for instance, to a radiation or
dust dominated ``power law" inflation, as well as to different classes of
de Sitter models (see also Ref. \cite{Prigogine}).
In addition, it is easy
to see that the energy density, for each
value of $\gamma$, scales
with the
temperature in the same fashion as happens for
equilibrium states. For
instance, in the era of radiation ($\gamma = 4/3$)
one obtains
$\rho = a T^4$. Only the time-dependence of the
quantities is modified (see section 5, specially Eq.(42)).

It should be observed that irreversible matter creation may also describe
the so-called super (or pole) inflationary expansion ($\dot H > 0$) in the
terminology of Ref \cite{Lucchin}. This kind of scenario appears, for
instance, in
string theory when taking into account the effects of the dilaton, a scalar
field capable of driving superinflation. Such a model has recently been
proposed as a possible alternative to standard inflation (see Ref.
\cite{Veneziano} and
references therein). In the present context, as one can see from (14), the
condition $\dot H > 0$ will be satisfied if $\gamma_{*} < 0$, that is,
$\psi > 3nH$.

Another important point is related with the end of inflation, that is,
the beginning of the FRW-type expansion. As the reader may conclude
himself, the condition for inflation to come to an end can be obtained
by the onset of violation of the above inequality (16). In fact, rewriting
(14) as

\begin{equation}
\frac {\ddot{R}}{R} = ( 1 -  \frac{3\gamma_{*}}{2}) H^{2}
\quad ,
\end{equation}
it is clear that the stage of accelerated expansion will finish
($\ddot R$ = 0) when
$\psi = (1-2/3\gamma)3nH$, that is, $\beta = 1 - 2/3\gamma$. For instance,
for $\gamma=4/3$ the universe evolves naturally from an exponential
inflation to a FRW-type expansion provided that the matter creation
decreases in the
interval $3nH/2 \leq \psi \leq 3nH$. For dust ($\gamma=1$), a slightly
different interval is required since the inflationary period will finish
when $\psi=nH$, that is, $\beta=1/3$. Parenthetically, we remark
that such conditions do not constrain the magnitude of the Hubble
parameter to assume any specific value. As we shall see in section 6,
this fact is closely related with the solution of the age problem which
plagues the standard model for all values of the curvature
parameter.

\section{The Simplest Class of Models}

Having in mind the choices previously made for $\psi$ we now propose
a specific matter creation scenario with a
slightly modified creation rate. As remarked in the introduction,
we are not interested here in presenting
a complete cosmological scenario with matter creation, that is, a model
describing the very early universe (including inflation) and the
late stages of the evolution.
Our goal is a much more limited one. We try to formulate a basic
scenario or equivalently, a kind of hot big-bang model minimally modified
due to the matter creation, over which the inflationary mechanism, as
discussed in the later section, or any other process,
may be further implemented.

In our opinion,
the simplest possible case, and probably the most physically appealing
too, at least for times later than the Planck era, is the one for which the
characteristic time scale for matter creation is the Hubble time itself.
Phenomenologically, this is equivalent (using (8)) to taking

\begin{equation}
\psi = 3 \beta n H \quad ,
\end{equation}
where now $\beta$ is a constant, which is presumably
given by the
particular physical model of matter creation.
The above creation rate also simplifies considerably the
task of solving eq. (13), since the effective ``adiabatic index"
 becomes $\gamma_{*}  =
\gamma (1-\beta) =$ const. In this case,  it is readily seen that
the generalized second-order
FRW equation for $R(t)$ given by (13) can be rewritten as

\begin{equation}
       R \ddot{R} + \Delta  \dot{R}^2 + \Delta k = 0     \quad,
\end{equation}

\noindent the first integral of which is

\begin{equation}
{\dot{R}}^2 =  {\frac {A}{{R}
^{2\Delta}}} - k
\quad ,
\end{equation}
\noindent where $\Delta = \frac{3\gamma(1-\beta)-2}{2}$ and
$A$ is a positive constant (see eq.(2)).

Using (20) one may express the energy density, the pressures ($p$ and
$p_c$) and the particle number density as functions solely of the scale
factor $R$ and of the $\beta$ parameter. In fact, inserting (20) into
(2), one obtains

\begin{equation}
\rho = \rho_o { ( \frac {R_o}{R} ) }^{3\gamma(1-\beta)}
\quad , \end{equation}
where $\rho_o = 3A/8 \pi GR_{o}^{3\gamma_{*}}$. The
above equation shows that the densities of
radiation and dust scale, respectively, as
$\rho_{r} \sim R^{-4(1 - \beta)}$ and
$\rho_{d} \sim R^{-3(1 - \beta)}$. Hence, in a model with radiation and matter,
 the transition from radiation to a dust dominated
phase, in the course of the expansion,
happens exactly as in the standard model.
Note also that, although formally defined by the same
expression (see (18)), the creation rates of radiation$(\gamma = 4/3)$ and
dust$(\gamma = 1)$ are not equal, as they seem to be at first sight.
For these cases, the above results are easily recovered in the
usual manner, e.g. defining $n = n_r + n_d$ and similar forms
for the energy density and pressure. The corresponding dominant
component then will determine the final
form of all physical quantities (see also comment below Eq. (24)).

Now, from (21), (5) and (6)

\begin{equation}
p_c = - \beta \gamma \rho_o
{ ( \frac {R_o}{R} ) }^{3\gamma(1-\beta)} \quad ,
\end{equation}
with the total pressure ${\it P}_t=p+p_c$ assuming
the form

\begin{equation}
{\it P}_t = (\gamma_{*} -1) \rho = [\gamma(1-\beta) -1 ] \rho_o
            {( \frac{R_o}{R} )}^{3\gamma(1-\beta)} \quad .
\end{equation}

Finally, by integrating (4) with $\psi$ given by (18), or more directly,
using eqs. (12) and (21), the expression for the particle number density
can be written as

\begin{equation}
n=n_o { ( \frac {R_o}{R} )}^{3(1-\beta)}
      \quad.
\end{equation}
A clarifying comment about the meaning of equations (21)-(24) is now in order.
Firstly, we note that (24) does not
depend explicitly on the $\gamma$ parameter. Thus, the same scale law
describes the evolution of the particle number density either for a
dominant or a nondominant component. The effect of matter creation in
both situations is measured by the $\beta$ parameter. The situation is
clearly different for the remaining equations, even though that those can
also be applied for the nondominant
component. In other words, for each expansion stage,
the explicit time dependent form of the energy density, equilibrium and
creation pressures as well as the scale factor (see Eq.(25)),
depends exclusively on the dominant component, however, the
creation of the
other component is not completely supressed. Of course, this is the same
kind of approximation commonly used in the standard FRW model. The only
difference is that even in the radiation era, the dust component
will have a nonvanishing creation pressure (see Eq.(22)), which although
negligible in comparison with the creation pressure of radiation, will be
responsible by the baryon production in that phase. Such considerations
will be important when we discuss the thermodynamic behavior of these models
(see section 5).

As expected, for $\beta=0$,  eqs. (18)-(24) reduce to
those of the
standard FRW model for all values of the parameters $k$ and $\gamma$ .
In this case, the unified solution of (19) or,
equivalently (20), was found by Assad and Lima \cite{A.L.}
in terms of
hypergeometric functions. Of course, such a solution can be adapted to the
present case simply by replacing the
``adiabatic index'' $\gamma$ by the
effective parameter $\gamma_{*}$.

For $k=0$, the solution of (19) for all values of $\gamma$ and $\beta$
can be written as

\begin{equation}
R= R_o { [ 1 + \frac{3\gamma (1-\beta )}{2} H_o (t-t_{o}) ] }^
   {\frac{2}{3\gamma (1-\beta )} } .
\end{equation}
Note that in the limit $\gamma \rightarrow 0$ the above solution
describes a de Sitter type universe
for any value of $\beta$. As remarked in the previous section, such a
solution can now
be obtained for $\beta \rightarrow 1$ and $\gamma$ assuming arbitrary values.
Further, for flat models with $\gamma_{*} > 0$, we can choose $t_{o} =
2 H_{o}^{-1} / 3\gamma (1-\beta )$, so that (25) assumes a more familiar form,
namely:
\begin{equation}
R(t) = R_o { [ \frac{3\gamma (1-\beta )}{2} H_o t ]}^
   {\frac{2}{3\gamma (1-\beta )} } \quad .
\end{equation}

\noindent If $\beta =0$, (25) and (26) reduce to the well known
expressions of the flat FRW model.

For $k\neq 0$, parametric solutions are usually more enlightening. By
introducing the conformal time coordinate,

\begin{equation}
dt = R d\eta \quad ,
\end{equation}
(19) can be recast as

\begin{equation}
R R'' + (\Delta -1 ) {R'}^{2} + \Delta k R^2 = 0 \quad ,
\end{equation}
where the primes denote conformal time derivatives.

Now, defining an auxiliary scale factor $z=R^{\Delta}$, it is readily seen
that (28) becomes

\begin{equation}
z'' = 0 \quad \quad  if \quad  \gamma = \frac{2}{3(1-\beta)}
\end{equation}
and
\begin{equation}
z'' + k {\Delta}^{2} z = 0 \quad if \quad \gamma \neq \frac{2}{3(1-\beta)}
\quad,
\end{equation}
whereas the first integral (20) is transformed into the energy conservation
equation:

\begin{equation}
\frac{1}{2} {z'}^{2} + \frac{1}{2} k {\omega}^{2} z^2 =
\frac{1}{2} {\omega}^{2} z_{*}^{2} \quad ,
\end{equation}
where $\omega = | \frac{3\gamma (1-\beta )-2}{2} |$
and $z_{*} = R_{*}^{2\Delta}$.

As one can see by direct substitution, the general
solution of (30) or
equivalently (31) is

\begin{equation}
z= z_{*}  \frac { \sin{  \sqrt{k} | \frac{3\gamma (1-\beta )-2}{2} |
(\eta + \delta)}} {\sqrt{k} } \quad ,
\end{equation}
where
$ \delta $ is an integration constant. Now, by choosing
$\delta = 0$ and using the inverse transformation
$ R=z^{1/ \Delta } $, the general
solution for the scale factor takes the following
form:

\begin{equation}
R(t)= R_{*} [ \frac { \sin{
      \sqrt{k} | \frac{3\gamma (1-\beta )-2}{2}| \eta}}
      {\sqrt{k}}  ] ^{\frac{2}{3\gamma(1-\beta) -2}} \quad , \end{equation}

\noindent and

\begin{equation}
t(\eta) = R_{*} \int [ \frac { \sin{  \sqrt{k} | \frac{3\gamma
          (1-\beta )-2}{2} | \eta}}
          {\sqrt{k} }]^{\frac{2}{3\gamma(1-\beta) -2}} \, d\eta + const.
\end{equation}

It should be remarked that the auxiliary scale factor $z=R^{\Delta}$
shows the same dynamic behavior appearing in the
standard FRW
model, namely it evolves as a free particle ($k=0$), a simple harmonic
oscillator ($k=1$)
or an ``anti-oscillator'' ($k=-1$). In this sense, some basic
characteristics of the standard FRW models are not modified, namely open
and flat universes expand forever whereas closed
geometries exhibit a turning
point either when the universe expands away from the singularity ($R=0$)
or starts contracting from $R=\infty$ with the models presenting a
big bounce. For completeness,
we observe that the physical meaning of $z_{*}$ or
equivalently $R_{*}$ is readily obtained from the first
integral (20). For instance, in the case of closed geometries with
$\gamma_{*} \neq 2/3$,
 $R_{*}$ is just the value
of $R$ at the turning point, that is, for which $\dot{R}(R_{*}) = 0$.
Accordingly, in the conformal time description we see from (31) that the
turning
point $z_{*}$ ($\gamma_{*} \neq 2/3$) corresponds to the
amplitude of the related spring-mass system of unit mass (SHO).
We leave it to the reader to verify that for singular flat models $R_{*} =
R_o$.

\vskip 0.3cm

\section{Thermodynamic Behavior}

The matter creation formulation considered here is a clear consequence of
nonequilibrium thermodynamics in the presence of gravitational fields
\cite{Prigogine,L.C.W.,C.L.W.}. In this context, unlike other
approaches to matter creation proposed
in the literature (Cf. , for instance, \cite{Hoyle} and \cite{Freese}),
the explicit thermodynamic
connection leads naturally to specific predictions on rates of variation
of the entropy per particle and of the temperature. As shown in Ref.
\cite{C.L.W.}, for the case of adiabatic matter creation, these rates are

\begin{equation}
\dot{\sigma} = 0 \quad ,
\end{equation}

\begin{equation}
\frac{\dot T}{T} = {( \frac{\partial p}{\partial \rho} )}_n
\frac{\dot n}{n} \quad ,
\end{equation}
where $\sigma$ is the dimensionless specific entropy and $T$ is the
temperature.

For pedagogical convenience we first discuss the time dependence of the
temperature. Using the
$\gamma-$law equation of state, the temperature evolution equation (36)
takes the form below,

\begin{equation}
\frac{\dot T}{T} = (\gamma -1) \frac{\dot n}{n} \quad ,
\end{equation}
the integral of which is

\begin{equation}
n(T) = n_o {( \frac{T}{T_o}
          ) }^{\frac{1}{\gamma-1}} \quad .
\end{equation}
Now, replacing into (38) the deduced relation between $\rho$ and $n$
given by (12), the former can be written as

\begin{equation}
\rho (T) = \rho_o {( \frac{T}{T_o}
          ) }^{\frac{\gamma}{\gamma-1}} \quad .
\end{equation}
The above expressions for $n(T)$ and $\rho (T)$, are exactly the general
expressions obeyed by a $\gamma$-fluid in the course of
adiabatic expansion \cite{L.S.}. As remarked in section 2,
for the case of radiation ($\gamma = 4/3$) the energy and particle
number densities scale, respectively, as $\rho_r \simeq T^4$ and
$n_r \simeq T^3$ so that  the created radiation necessarily satisfies
the usual equilibrium relations. This is a remarkable result. Differently
from other approaches for matter creation, where expressions like (38) and
(39) need to be assumed (see, for instance, Refs. \cite{Freese}), the
equilibrium
relations here follow as a  consequence of the ``adiabatic'' condition. In
fact, as
discussed
in detail in the literature \cite{C.L.W.,L.G.}, the condition (35) determines
simultaneously the creation pressure form (5) and the temperature
equation given by (36).
On the other hand, combining $\rho (R)$ given by (21),
with (42), we obtain the following temperature law :

\begin{equation}
T = T_o {( \frac{R_o}{R} ) }^{3(\gamma-1)(1-\beta)}
\quad .
\end{equation}
The above result shows us that exponential inflation
($\beta=1$) occurs isothermically regardless of the value of $\gamma$(see
section 2). In fact,
generically, the $\beta$ parameter works in the opposite sense of
the expansion, that is, diminishing the cooling rate with respect to the
case with no matter creation. In particular, instead of the usual
result, $RT=$ const., valid for
radiation($\gamma = \frac{4}{3}$), we find
$TR^{1-\beta}=$ const. Note also that by integrating (37) with
$n=\frac {N}{R^{3}}$, the temperature law assumes a
new form where the
$\beta$ parameter does not appear explicitly, namely:
$N^{(1-\gamma)}T{R^{3(\gamma - 1)}}=$ const., which makes transparent
the conclusion that if $N=$ const., the usual evolution law is
recovered. This formula does not depend on the specific creation rate
assumed in the present paper. In particular, for $\gamma=4/3$ one has

\begin{equation}
N^{-1/3}TR = const.  \quad,
\end{equation}
as one should expect(see Ref. \cite{C.L.W.}). Now, recalling that for
FRW geometries the frequency redshifts obeying, $\nu \sim R^{-1}$, the above
temperature law leads inevitably to the conclusion that the usual Planckian
spectrum is destroyed in the course of the evolution, in particular, after
decoupling. Nevertheless,
as recently shown by one of us\cite{JL 95}, for ``adiabatic'' matter creation
the preserved spectral distribution is given by

\begin{equation}
\label{eq:forend3}
\rho _{T}(\nu) = {(\frac {N(t)}{N_o})^\frac {4}{3}} \frac {8 \pi h}{c^{3}}
\frac {\nu ^{3}}
{exp[(\frac {N(t)}{N_o})^\frac{1}{3} {\frac {h\nu}{kT}}]  - 1}   \quad,
\end{equation}
where $N$(t) is the comoving time dependent number of photons and $N_o$ is the
constant value of $N$ evaluated at some fixed epoch, say, the present time. The
above distribution is a
consequence of the temperature evolution law as given
by (41).
When there is no creation, $N(t)=N_o$, and the usual Planckian form is
recovered. In addition, it is readily seen that the equilibrium relations are
recovered using such
a spectrum. In fact, for $\gamma=\frac {4}{3}$, it follows from (38) and (39)
that $n \sim \rho^{\frac {3}{4}}$, and
by introducing a new variable
$x=(\frac {N}{N_o})^\frac {1}{3} \frac {h\nu}{kT}$,  it is easy to see that

\begin{equation} \label{eq:rOCONST}
\rho (T) =\int_{0}^{\infty} \rho_{T}(\nu )d\nu =
aT^{4} \quad,
\end{equation}
where $a$ is the usual radiation density constant.
The spectrum given by (42) seems to be the most natural generalization of
Planck's radiation formula
in the presence of ``adiabatic'' photon production.
More important still, (42)
cannot be distinguished from the usual
blackbody spectrum at the present epoch when we take
$T=T_{o}$ and $N(t_{o})=N_{o}$. Therefore,
models with ``adiabatic'' photon creation may be compatible with
the present isotropy and spectral distribution of the microwave background.
Of course, since photons are injected satisfying
(42), which is preserved in the course of the evolution, there will be no
distortions in the present
relic radiation spectrum. Note that (42) is preserved precisely due to
validity of the temperature law (41). In this concern, the macroscopic
formulation adopted here seems to be naturally connected with some
fundamental cosmological irreversible mechanism
(based on microphysics), in which photons are quantum mechanically produced
with the above thermal spectrum and baryons are asymmetrically created\cite{CO
2}.

Let us now consider the entropy behavior as defined in (35). Since $\sigma =
S/N$, where $S$ and $N$ are, respectively, the entropy of the dominant
component and the corresponding number of particles, (35) can be
rewritten as

\begin{equation}
\frac{\dot S}{S} = \frac{\dot N}{N} \quad .
\end{equation}
Hence, due to the matter creation processes, the universe does not expand
adiabatically as happens in the standard
FRW models\cite{Com}. Besides, since
up to a constant factor one has $N=nR^3$, it
follows from (24) that $N$ increases as
a power of $R$, that is,

\begin{equation}
N=N_o {( \frac{R}{R_o} )}^{3\beta} \quad .
\end{equation}

Further, from eq. (37), $S=S_o (N/N_o)$, and using the above
expression one may write the photon entropy as (from now on indexes r and b
refer, respectively, to radiation and baryon component(dust))

\begin{equation}
S_{r}= S_{or} {( \frac{R}{R_o} ) }^{3\beta}
\quad ,
\end{equation}
where $S_{or} \approx 10^{87}$ is the present observed radiation entropy
(dimensionless). As happens in the standard model,
although remaining nearly constant, the specific entropy defined by (35) does
not play an important physical role. As we know, some physically meaningfull
informations, as for instance, in nucleosynthesis studies as well as for
the structure formation problem, are encoded in the specific radiation
entropy per baryon. Such a quantity, defined by
$\sigma_{rb} = \frac{S_r}{N_b}$, is proportional to the photon-baryon
ratio and, up to short aniquillation period, also remains constant in
the standard model.
In the present context, since photons are thermally produced and baryons
are  continuously created, a nearly constant behavior of $\sigma_{rb}$
should also be expected. In fact, the net number of baryons in the comoving
volume is
given by(see discussion below Eq.(24))

\begin{equation}
N_{b}=N_{ob} {( \frac{R}{R_o} )}^{3\beta}
\quad ,
\end{equation}
where $N_{ob}$ is the  present baryon number. It
thus follows from (46) that
$\sigma_{rb} = \frac {S_{or}}{N_{ob}}
= \sigma_{o} \approx 10^{10}$, is the present photon to baryon ratio. As a kind
of consistency check, we notice that if one write the specific entropy(per
baryon) in the usual form

\begin{eqnarray}
\sigma_{rb}= \frac {4a{T_{r}}^3}{3n_{b}}
\quad ,
\end{eqnarray}
the above result is recovered since $T_{r}^{3}$
and $n_{b}$ evolve following the same scale law(see Eqs.(40) and (24)).
Therefore,
the adiabatic formulation does not provide any explanation for the
present value of photon to baryon ratio. As happens
in the big-bang
model, the value of $\sigma_o$ is just an initial condition.
In the present model, particles are created in spacetime with the same
temperature as the
already existing ones have, otherwise the specific entropy could not
remain constant, as happens in the more general formulation proposed in
Ref \cite{C.L.W.}. Naturally,
from a thermodynamic point of view the model is irreversible. The burst
of entropy is closely related with the creation of matter and radiation.

We would like to stress that all important thermodynamic results of the
standard
FRW models like $S=S_o$, $N=N_o$ and the radiation temperature
scaling  $T \propto R^{-1}$ are recovered for $\beta=0$. Finally,
it is interesting to remark that the models
presented here may significantly alter the standard predictions
of cosmic abundances, since they alter
the expansion rate and predict a new temperature law. Such
results points to possible limitations
on the $\beta$ parameter imposed through constraints from
nucleosynthesis. This issue will be addressed elsewhere.

\vskip 0.3cm
\section{Some Observational Aspects}

Now we illustrate some observable predictions of the models proposed in
the preceding sections. Following standard lines we define
the physical parameters  $q = - {\frac{R\ddot R}{{\dot R}^2}}$
(deacceleration parameter), $H=\frac{\dot R}{R}$ (Hubble parameter)
and $\Omega = \frac{\rho}{\rho_c}$ (density parameter), where
$\rho_c = \frac{3 H^2}{8 \pi G}$ is the critical density.

Inserting the above quantities into Eqs. (2) and (19) we have

\begin{equation}
\Omega = \frac{2q}{3\gamma(1-\beta) -2}
\end{equation}
and
\begin{equation}
\frac{k}{R^2} = (\Omega -1)H^2 \quad .
\end{equation}
Therefore, it is clear that if $\Omega >1$, that is, if
$q > \frac{3\gamma(1-\beta) -2}{2}$ the
universe is positively curved with $\rho > \rho_c$, whereas
if $q \leq \frac{3\gamma(1-\beta) -2}{2}$ it is negatively curved or flat,
respectively, with $\rho \leq \rho_c$. For $\beta=0$ the usual expression
for FRW models are recovered. However, the positivity
of $\Omega$ does not restrict q to be positive if $\gamma > 2/3$ as happens
in the standard universe. To be more specific, in the present dust phase
($\gamma = 1$), the above expressions reduce to

\begin{equation}
\Omega_o = \frac{2q_o}{1-3\beta} \quad ,
\end{equation}

\begin{equation}
\frac{k}{{R_o}^2} = (\Omega_o -1){H_o}^2 \quad .
\end{equation}
Therefore, if $q_o = \frac {1-3\beta}{2}$ we have $\Omega_{o}= 1$ and from
(52), the presently observed universe is flat. However, regardless of the value
of $k$, the
deacceleration parameter may be negative since the constraint $\Omega_o > 0$
can be satisfied for $q_o < 0$, provided that $\beta > 1/3$. As we shall see
next, such a fact allows us to solve the age problem in the present context.

As we know, the age of the universe is found by integrating the generalized
first integral (20). By expressing the constant A in terms of $\Omega_o$,
$R_o$ and $H_o$ it is straightforward to show that

\begin{equation}
{ ( \frac {\dot {R}}{{R_{o}}} ) }^2= H_o^2 [ 1 - \Omega_{o}
+ \Omega_{o} { ( \frac{R_{o}}{R} ) }^{2 \Delta} ] \quad ,
\end{equation}
the solution of which may be expressed as a formula for the time t in
terms of R,

\begin{equation}
t - t_{*} = {H_o}^{-1} \int_{{R_{*}}/R_o}^{R/{R_o}}
            { [ 1-\Omega_o + \Omega_o x^{-2 \Delta} ] ^{-1/2}  dx } \quad ,
\end{equation}
where $t_{*}$ is the time for which $R=R_{*}$.

For singular models, the present age of the universe is defined by
taking $R_{*}=t(R_{*})=0$ so that it is given by

\begin{equation}
t  = {H_o}^{-1} \int_{0}^{1}
     { [ 1-\Omega_o + \Omega_o x^{-2 \Delta} ] ^{-1/2} dx } \quad ,
\end{equation}
which for $\beta = 0$, that is, $\Delta = \frac{3\gamma-2}{2}$, is exactly
the same as in the standard model. In Figs. 1 and 2 we show the age
of the universe in units of $H_o^{-1}$ for the cases of dust and radiation
dominated universes and some selected values of $\beta$. Observe
that the
universe can be old enough even when considering a radiation dominated phase
today, as has been suggested sometimes. Another important conclusion is that
if $\beta \geq 1-2/3\gamma$ and $0 \leq \Omega_{o} \leq 1$ the oldest
universe is the flat one ($\Omega_o = 1$). In this context, our matter creation
ansatz (18) changes the predictions of standard cosmology in such a way
that it solves the problem of reconciling observations with the inflationary
scenario. Note also that open models with a small parameter are also ruled
out by recent data regardless of the value of $\beta$.

The solution of the age problem and its noticeable dependence on the $\beta$
parameter can be exactly determined for flat models. In this case
($\Omega_o =1$), one can see that the age parameter reduces to

\begin{equation}
H_o t_o = \frac{2}{3 \gamma (1-\beta) } \quad .
\end{equation}

\noindent In a matter-dominated universe we have then
$H_o t_o = 2/{3(1-\beta)}$, and in this form it is easiest to see how
matter creation could solve the age problem suggested by the latest
direct measurements of the Hubble constant done by Pierce {\it
et al.}\cite{Pierce} and Freedman {\it et al.}\cite{Freedman} who found,
respectively, $H_o = 87 \pm 7 Km s^{-1} Mpc^{-1} $ and
$H_o = 80 \pm 17 km s^{-1} Mpc^{-1}. $ Assuming no matter creation
($\beta = 0$), these values of $H_o$ imply that the expansion age of a
dust-filled, flat universe would be about either  $7.3 \times 10^{9}$ years or
$8.2 \times 10^{9}$ years, in
direct contrast to the measured ages of some stars and stellar systems,
believed to be at least some $(16 \pm 3)\times 10^{9}$ years
old or even older if one adds a realistic
incubation time \cite{Idade}. Such measurements restrict the
parameters $H_o t_o$ to
the following intervals (P and F reffer to the values of $H_{o}$
given, respectively, in Refs.\cite{Pierce} and \cite{Freedman})

\begin{equation}
1.09 \leq H_{o} t_{o} \leq 1.86  \quad  \quad ({\bf P})
\end{equation}
and

\begin{equation}
0.85 \leq H_{o} t_{o} \leq 1.91  \quad  \quad ({\bf F}),
\end{equation}
when in the standard flat model ($\gamma=1, \beta=0$), one would
obtain exactly $2/3$. In this fashion, these recent
measurements point to a serious crisis of
the standard model. It would not be surprising if
expectations that it could, at least marginally, be
compatible with the age of the universe, will
gradually be forgotten (see Figs.1 and 2).

As can be easily seen from (56), matter creation naturally solves
this problem, increasing the parameter $H_{o} t_{o}$ while preserving the
overall
FRW evolution scheme. From (56), (57) and (58),  the constraints on the
$\beta$ parameter for both cases are readily computed to be

\begin{equation}
0.38 \leq \beta \leq 0.64 \quad \quad ({\bf P})
\end{equation}
and
\begin{equation}
0.21 \leq \beta \leq 0.64 \quad \quad ({\bf F}).
\end{equation}
Note that due to the error bar in the values of $H_o$, the upper bound of
$\beta$ does not depends on the particular set of measurements. Naturally,
a more realistic range of $\beta$ probably will have an upper bound
slightly smaller than 0.64 and a lower bound between 0.21 and 0.38.
This will happen, for instance, if the future measurements of $H_{o}$
improve by at least one order of magnitude. Parenthetically, there are
some independent indications requiring the present expansion rate to lie in
the range $H_o = 80 \pm 5 Km s^{-1} Mpc^{-1}$ (see \cite{Turner1} and
references there in). With such a precision, the  $\beta$ parameter will
fall on the interval $0.34 \leq \beta \leq 0.60.$

As remarked earlier, the solution of the age problem is due to the
possibility of $q_o < 0$ nowadays. For instance, from (49) with $\Omega_o=1$ we
can rewrite the result (59) as $H_o t_o = 1/(1+q_o)$. For $\beta$ in
the above interval, we have for a dust filled
universe $-0.44 \leq q_o \leq -0.1$. In connection to this we note that
cosmological constant models with a reasonable value of $\Omega_\Lambda$
are also believed to solve easily the ``age problem''. However, as recently
shown by Maoz and Rix \cite{Maoz}, the computed rate of gravitational
lensing in such models constrain severely $\Omega_\Lambda$ when confronted
with the existing lensing observations. Analogously, such a result seems
to point out to similar limits on the $\beta$ parameter, thereby leading
to values of $H_o t_o$ below the range given by (57) or (58). In this concern,
we
remark that models with matter creation behave like scenarios driven by
a decaying $\Lambda term$, instead of models with cosmological constant
\cite{Freese,CarvalhoLW}. Moreover, for this kind of models, a lower lensing
rate is
usually predicted since the distance to an object with redshift $z$ tends
to be smaller than the distance to the same object for a model with $\Lambda$
constant \cite{T.W.}.

As we have discussed in the previous section, photons are always created
in equilibrium with the already existing radiation. However, as
we have seen, they are not dominant nowadays.
In the present dust phase, the matter
creation rate is given by

\begin{equation}
\psi_o = 3 \beta n_o H_o \quad .
\end{equation}

For $n_o \sim 10^{-6}$ nucleons $cm^{-3}$ and $H_o^{-1} \sim 10^{10} yr$
we have $\psi_o \sim 10^{-16}$ nucleons $cm^{-3} yr^{-1}$, which is nearly
the same rate predicted by the steady-state universe
\cite{Narlikar,Hoyle} and also by some
decaying $\Lambda$ cosmologies\cite{CarvalhoLW}. Of course, this matter
creation rate is presently
far below detectable limits.

\section{Conclusion}

Almost all research efforts related to the physical
processes in the cosmological domain have been
devoted to the standard model in its different
phases. As we know, such a model is thermodynamically
characterized
by two different although related features, namely: entropy
conservation ($S^{\alpha}_{;\alpha}=0$) and number particle conservation
($N^{\alpha}_{;\alpha}=0$), where $S^{\alpha}$ and $N^{\alpha}$ are,
respectively, the four-vectors of entropy and number of particles.

In this paper we have investigated some cosmological consequences
arising when one changes the second (and consequently the first) of the
above properties. Of
course this is not new as may easily be observed in the extensive
literature on this subject
(see \cite{Narlikar} and references therein). The new fact
justifying the present work is that we have considered
a recent thermodynamic approach for which matter
creation, at the expense of the gravitational field, has
been properly constrained by the usual requirements of
nonequilibrium thermodynamics [22-25]. In this
context, thermodynamic
predictions such as the temperature law and the variation rate of the entropy
are computed from first principles. Such an approach points to
a possible revival of interest in models with radiation and/or matter
creation, which is one of the most challenging problems of theoretical
physics.

For ``adiabatic"  matter creation a general
expression relating the particle number and energy densities which holds
regardless of the creation
rate has been deduced and the conditions for generating
inflation have also been established. It was also shown
that a minimally modified big bang model with adiabatic
matter creation can be easily implemented and predicts
interesting cosmological consequences. In fact,
for a creation rate $\psi = 3 \beta n H$, the known
perfect fluid solutions can be adapted by using a
slightly modified ``adiabatic index"
$\gamma_{*} = \gamma (1-\beta) $. The thermodynamic
behavior is also readily computed making clear
how the above mentioned properties of the standard model
are quantitatively changed. In particular, instead
of constant entropy and
$T \sim R^{-1}$, we found  $T \sim R^{-(1-\beta)}$ and
$S \sim R^{3\beta}$. In addition, since the specific entropy of radiation
is very high, one can show that as long as matter is in thermal contact
with radiation it will follow the same temperature law as the radiation,
so that the thermal history is accessible as in the
standard model.
The model is also able to harmonize
a FRW-type picture with the discrepancy existing
between the latest measurements of the Hubble parameter
and the age of the universe, as predicted by the
standard model. Of course, in order to have a viable
alternative to the standard FRW model, other well
known cosmological tests need to be investigated.
Further details of our model will be published elsewhere.

\section{Acknowledgments}
It is a pleasure to thank R. Brandenberger and A.
Sornborger for many valuable suggestions and a critical reading
of the manuscript. Many thanks are due also to R. Moessner and
M. Trodden for the permanent stimulus and interest in this work.
One of us (JASL) is grateful for the hospitality of the Physics
Department of Brown University. This work was partially
supported by the Conselho Nacional de Desenvolvimento Cient\'{\i}fico e
Tecnol\'ogico - CNPq (Brazilian Research Agency), and by the US
Department Of Energy under grant DE-FG02-91ER40688, Task A.

\newpage
\begin{center}
{\bf Captions for Figures
\vskip 1cm
Fig. 1}
\vskip 0.5cm
\end{center}
\it
\begin{quote}
{\small
{\bf Fig. 1} - The age parameter of a matter dominated universe as
a function of $\Omega_o$ for some selected values of $\beta$. The
horizontal dotted and solid lines indicate the latest observational constraints
(see Eqs. (52) and (53)). The solid curves represent two limiting cases, the
standard model ($\beta=0$) and a de Sitter type universe
($\beta=1$). Curves A, B and C respectively represent models with
$\beta = 0.44$, $\beta=0.54$ and $\beta=0.63$. According to Pierce et al. data,
the standard dust model is ruled out regardless of the value of $\Omega_o$
whereas for Friedman et al. only extremely open FRW models
may be compatible with the observations. For $\beta > 1/3$ the matter creation
process
rehabilitates the flat universe, as predicted by inflation.}

\end{quote}

\vskip 1cm

\begin{center}

{\bf Fig. 2}

\end{center}

\vskip 0.5cm

\begin{quote}
{\small
{\bf Fig. 2} - The same graph of Fig. 1 for a radiation dominated
universe. The solid curves now indicate the radiation filled FRW
model ($\beta=0$) and a de Sitter-like radiation universe ($\beta=1$).
Generically, the curves are displaced downwards in
comparison with the case of dust so that the models
are compatible with data only for higher values of $\beta$.}
\end{quote}


\begin{thebibliography}{50}
\bibitem{CO 2} As discussed in Ref.\cite{JL 95}, there is a crucial test for
this kind of cosmologies. From (41) one can see that the temperature
redshift relation is given by $T = T_o(1 + z)(\frac {N}{N_o})^{\frac{1}{3}}$.
Thus, universes with
``adiabatic`` matter creation are, for a fixed value of
$z$, cooler than the standard model. Such a relation may be verified observing
the atomic or molecular transitions in absorbing clouds at high
redshifts\cite{SO 94}.
\bibitem{Com} Note that Eq.(44) can be rewritten as $\frac{\dot S}{S}=
\frac{\psi}{n}$. It thus follows from thermodynamic second law that
$\psi \leq 0$, that is, the spacetime can only create matter,
the reverse process being thermodynamically forbidden\cite{Prigogine}. In our
case, since $\psi= {\beta}{nH}$, we see
that models with $H<0$ and positive values of $\beta$ are ruled out from a
thermodynamic point of view. For closed models with a turning point, this means
that the $\beta$ parameter must change sign when the universe radius increases
beyond its maximum value.
\end{thebibliography}
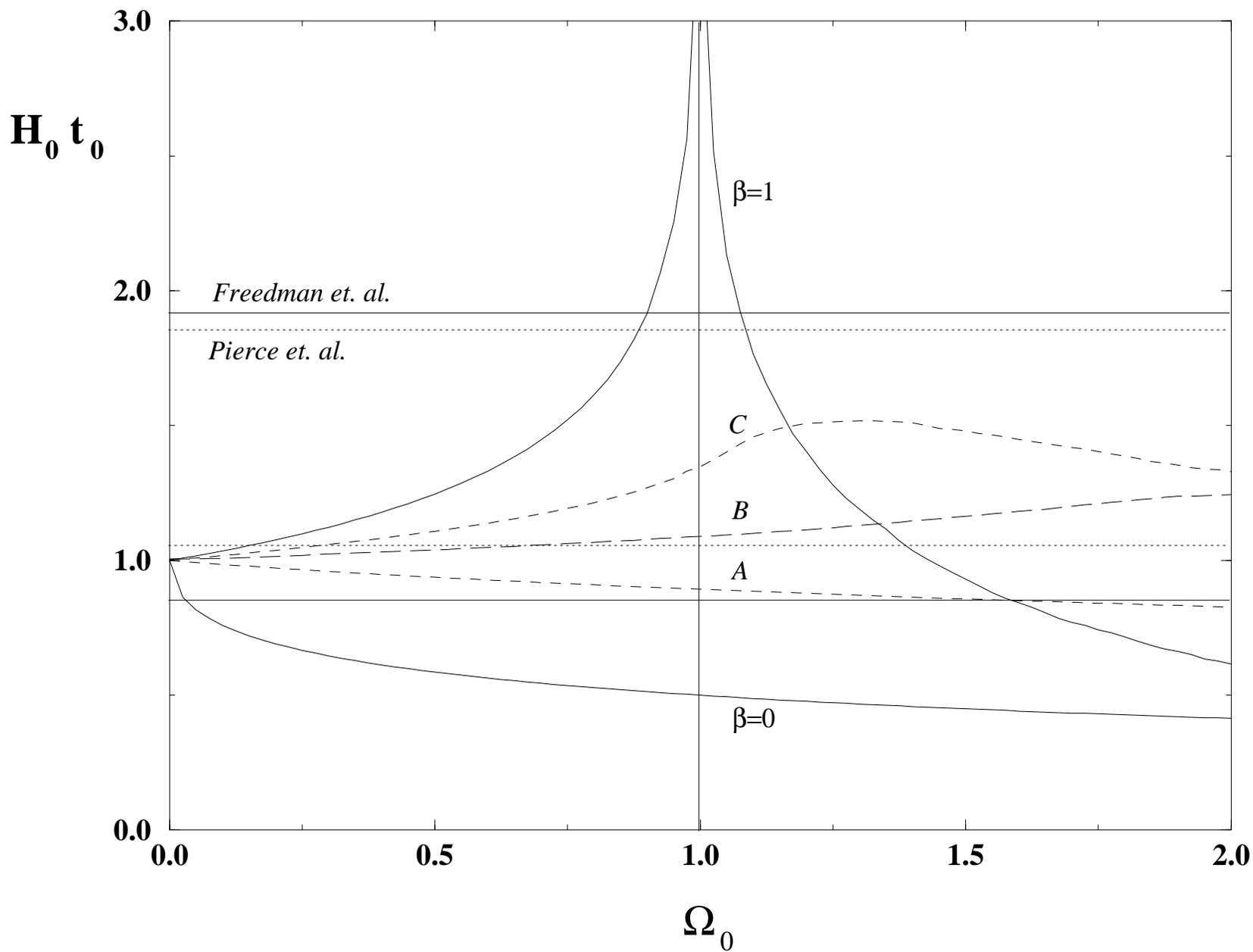
\end{document}